\documentclass[12pt,a4paper]{article}
\usepackage{graphicx}% Include figure files
\usepackage{dcolumn}% Align table columns on decimal point
\usepackage{bm}% bold math
\usepackage{xcolor}
\def\bea{\begin{eqnarray}}
\def\eea{\end{eqnarray}}

\newcommand{\nn}{\nonumber}

\def\beq{\begin{equation}}
\def\eeq{\end{equation}}

\usepackage{amssymb}
\usepackage{amsmath}

\begin{document}

%\preprint{APS/123-QED}

%\vspace{14mm}
\begin{center}
       { \Large\bf
Aspects of $(d+D)$-dimensional Anisotropic Conformal Gravity}
\end{center}

%\author{Joohan Lee}
%\email{joohan@kerr.uos.ac.kr} \affiliation{Department of Physics,
%University of Seoul, Seoul 130-743 Korea}
%\author{Tae Hoon Lee}\email{thlee@ssu.ac.kr}
%\affiliation{Department of Physics and Institute of Natural
%Sciences,\\ Soongsil University, Seoul 156-743 Korea}
\vspace{3mm}
\begin{center}
\large
Jae-Hyuk Oh$^1$ and Phillial Oh$^2$
\end{center}

$^1${\it Department of Physics, Hanyang University, Seoul 04763, Korea}
%\begin{center}
%and 
%\end{center}

$^2${\it Department of Physics
and Institute of Basic Science, Sungkyunkwan University, Suwon
16419, Korea}

\begin{center}
{\tt jaehyukoh@hanyang.ac.kr, ploh@skku.edu}
\end{center}
%\date{January 22, 2003}%
\date{\today}
\vspace{3mm}
\begin{abstract}
\noindent We discuss various aspects of anisotropic gravity in $(d+D)$-dimensional spacetime where $D$ dimensions are treated as extra dimensions. It is based on the foliation preserving diffeomorphism invariance and  anisotropic conformal invariance. The anisotropy  is embodied by introducing a factor $z$ which discriminates the scaling degree of the extra $D$ dimensions against the $d$-dimensional base spacetime and  Weyl scalar field which mediates the anisotropic scaling symmetry. There is no intrinsic scale but a physical scale $M_*$ emerges as a consequence of spontaneous conformal symmetry breaking.
Some vacuum solutions are obtained and we discuss an issue of `size separation' between the base spacetime and the extra dimensions. The size separation means large hierarchy between the scales appearing in the base spacetime and the extra dimensions respectively. We also discuss interesting theories obtained from our model. 
In the case of (4,1), we propose a resolution of hierarchy problem and discuss comparison with the results of the brane-world model.
In a $(d,D)=(2,2)$ case, we suggest a UV-complete unitary quantum gravity which might become Einstein gravity in IR.  In a certain (2,1) case, we obtain CGHS-model.
\end{abstract}

%\pacs{ 95.35.+d, 04.20.Jb, 95.36.+x}
%\keywords{?cosmology; higher dimensional gravity; phantom field;
%coincidence problem}
%\preprint{hep-th/yymmnn}

%\maketitle

%Contents
\newpage
\section{Introduction}

One of the main motivations to consider spacetime dimensions other than four in theoretical physics is to tackle with the quantum gravity. In string theory which is regarded as the most promising theory of quantum gravity\cite{Johnson:2000ch,bbs}, the spacetime is enlarged to accommodate extra dimensions at short distances. The higher-dimensional gravity theories like Kaluza-Klein theory\cite{Appelquist:1987nr} or the brane-world models\cite{Raychaudhuri:2016kth} are also based on the assumption of extra dimensions.
The reverse path of reducing the spacetime dimensionality to two at high energy also has been studied\cite{Carlip:2019onx}. 

In the higher dimensional  theories, the dimensions are scale-dependent.
At small enough distances, one can probe %the spacetime involves 
extra dimensions. However, at a distance larger than the scale, the spacetime becomes effectively four-dimensional, the (compact) extra dimensions are no longer being observed.
Under the circumstance that the exact nature of spacetime is still far from being accessible, it is worthwhile to consider diverse possibilities about spacetime and one of the natural questions is to ask about isotropy of spacetime.
Is spacetime isotropic in all its directions? The current Universe favors this and it is well accepted that the four-dimensional spacetime we are living in is uniform in all directions\cite{Long:2002wn,Joyce:2014kja}\footnote{Abandoning the Lorentz invariance and equal-footing treatment of time and space in UV is advocated in Ref.\cite{Horava:2009uw} for quantum gravity.}. However as far as extra dimensions are concerned, it leaves us with a possibility that they might not share the
same properties as those of the four-dimensional spacetime.
Therefore, it is worthwhile to undertake the task of investigating whether the higher-dimensional spacetime has uniform physical properties in all directions.

Recently, there is an attempt to construct a  higher-dimensional gravity theory where the four-dimensional spacetime and the extra dimensions are not treated on an equal footing\cite{Moon:2017rox}. The essential feature is given by an introduction of a real parameter $z$ which measures the variance of the anisotropy between the four-dimensional spacetime and the extra dimensions, and it makes the theory enjoy the anisotropic conformal invariance.
There are no restrictions on the value of $z$ at the classical level. However, the Planck and WMAP's cosmic microwave background data\cite{Hinshaw:2012aka,Ade:2013zuv} can restrict its value. A cosmological test in a five-dimensional theory\cite{Kouwn:2017qet}  was presented in order to check which specific value of $z$ is preferred by applying the data to the theory.

%Put the references here
%\textcolor{blue}
{Such an anisotropic scaling appears in the other fields of high energy physics too. In the Lifshitz fixed point, it is well known that time and space scale differently as $t\rightarrow b^zt$ and $\vec x \rightarrow b \vec x$, where $b$ is a positive real number. Especially, in AdS/CFT context, there is much of discussion about construction of dual gravity models\cite{Kachru:2008yh} and black hole(blackbrane) solutions\cite{Danielsson:2009gi} as a dual description of the Lifshitz fixed point. The methodology of AdS/CFT applies to holographic condensed matter and super fluidity\cite{Park:2016wch,Hartnoll:2009sz,Herzog:2009xv,Horowitz:2009ij}, which respect Lifshitz scaling. On top of these, there is discussion of  anomalies of anisotropic scaling symmetry in quantum level\cite{Arav:2014goa}. There is also an application of geometry enjoying anisotropic scaling symmetry to cosmology\cite{Kiritsis:2009sh}, where the authors discuss the possibility of escaping initial singularity problem.}

In this work, we generalize such an anisotropic scaling symmetry to $(d+D)$-dimensional spacetime where $D$-spatial dimensions are being treated as the extra dimension.  The extension provides opportunity to have discussions not only for $d=4$ theory but also for other interesting cases of $d$ like $d=2$ and $d=3$ among others. 
     The generalization  is straightforward and based on two compatible
symmetries of the foliation preserving diffeomorphism (FPD) and the anisotropic conformal transformation. The anisotropy is first implemented in $(d,D)$-ADM decomposition by keeping general covariance only for the  $d$-dimensional spacetime. This can be achieved by adopting FPD in which the foliation is adapted along the extra dimensions. The theory is extended to
conformal gravity by
introduction of a Weyl scalar field, $\phi$. The scalar field is introduced as a Stueckelberg field of a scale transformation of metric, where the base spacetime and the extra dimensions transform in different 
ways\footnote{%\textcolor{blue}
{We note that there is a similar attempt with our work\cite{Perez-Nadal:2016tzr}. In this paper, the authors consider a scalar field theory as Landau-Ginzberg theory of scalar order parameters near the Lifshitz fixed point. They study such a theory in product space given by $\mathcal M_d\times \mathcal N_D$, where $\mathcal M_d$ and $\mathcal N_D$ are $d$- and $D$-dimensional Euclidean manifolds. The theory enjoys a local anisotropic scaling symmetry where  each metric defined on $\mathcal M_d$ and $ \mathcal N_D$ scales differently as well as is invariant under a product of diffeomorphism groups in each space of $\mathcal M_d$ and $\mathcal N_D$, given by $Diff(\mathcal M_d)\times Diff(\mathcal N_D)$. }}.

%\textcolor{blue}
{Our main claims are threefold. Firstly, our model can resolve hierarchy problem by employing a rather mild control of the vacuum expectation value (vev) of Weyl scalar  at $\phi=\phi_0$ and the parameter $z$ with an assumption of %symmetry breaking of the Weyl scalar field. 
anisotropic conformal symmetry breaking.
Secondly, the model will provide interesting lower-dimensional theories, which become unitary when the parameters in the theory are chosen properly. Finally, Our theory is also able to cause large distinction of scales between  $d$-dimensional spacetime and $D$-dimensional space by a similar mild adjustment of the scalar vev and $z$. These features will be explained below and the details will be given in Sec.\ref{Scale Emergence and Large Hierarchy}, Sec.\ref{Lower Dimensional Gravity}, and Sec.\ref{Vacuum Solution} in order.}

%\textcolor{blue}
{Firstly, we suggest a possible scenario for a resolution of the hierarchy problem and discuss comparison with the Randall-Sundrum(RS) model\cite{Randall:1999vf,Randall:1999ee}. 
We assume that  spontaneous conformal symmetry breaking   brings in and by which a mass scale $M_*$ appears, even if we starts with no scales.
It turns out that our model can provide a large scale distinction between the original mass scale, $M_*$ given in certain $(d+D)$-dimensional field theories and a physical mass scale in their effective $d$-dimensional theories. To illustrate this, we consider a massive scalar (matter) fields and its mass scale is given by $M_*$ since there is no other scales in the $(d+D) $-dimensional theory. However, its effective mass defined in $d$-dimensional theory after integrating out extra $D$ dimensions can show a large hierarchy from its original mass scale, $M_*$.
The ratio of the two scales depends on the parameter $z$ and so a tuning of  $z$ is needed to fit the phenomenological results. 
However, we stress that the well-studied models like RS model\cite{Randall:1999vf,Randall:1999ee} also needs a tuning of parameters in the theory to fit the phenomenological data.
In the RS model, they suggest a space with $M_4\times S^1$ with two different 3-branes put at certain locations on $S^1$(visible and hidden branes), where $M_4$ is the Minkowski  spacetime and the $S^1$ is a circle being an extra dimension. The ratio of the  mass scale on the hidden brane and the effective mass scale on the visible brane in the RS model depends on the circle size of the extra dimension, the size of $S^1$. The role of parameter $z$ is similar to the circle size in the RS model.}

Secondly, we discuss lower-dimensional theories derived from our model.
Our model possesses an anisotropic scale invariance, and so the action does not contain any dimensionful parameters.  %\textcolor{blue}
{However, we assume that the Weyl scalar field, $\phi$ can have a vev, $\phi_0$ by an appropriate symmetry breaking mechanism %thanks to conformal invariance 
and  the theory exhibits some interesting properties due to the scaling parameter  $z$.
For a specific example, in $d=D=2$ case, when $z$ is a negative and small number as
$-0.1\leq z<0,$ together with order one value of the Weyl scalar vev, $\phi_0$,
the effective theory reduces to two-dimensional conformal gravity theory coupled with nonlinear sigma model, where the target scalar fields interacting with gravity are furnished by the metric components of the extra dimensions. We note that this feature is quite general and can apply such an argument to generic $d$ and $D$ cases.} 
%This can be achieved with a suitable gauge fixing of the Weyl scalar field to a constant. 
All the other interactions coming from  the metric components involving the shift vector vanish in this limit regardless of the  value of $D.$   The derivative along the extra dimensions all disappears  and the extra dimensions become virtually `deactivated' in this limit. This effective two-dimensional gravity theory can be renormalizable and unitary with suitable choice of the parameters in the Lagrangian.

% starting from  the absence of  dimensionful parameter. 
%\textcolor{blue}
{Finally, we discuss vacuum solutions and another hierarchy of scales between the $d$-dimensional spacetime and the extra $D$-dimensional space appearing in the solutions in our model.
% Then,  depending on the value of $z$,  there could be a large separation between the physical mass parameters and $M_*$. 
We get a vacuum solution by assuming that the $(d+D)$-dimensional metric is a product form as $\mathcal G_d \times \mathcal G_D$, where $\mathcal G_d$ is the metric of $d$-dimensional spacetime and $\mathcal G_D$ is the metric of $D$-dimensional space respectively. The spacetime,
$\mathcal G_d$ or $\mathcal G_D$ are either de-Sitter or anti de-Sitter space, where their effective cosmological constants are given by $\Lambda^{(d)}$ and $\Lambda^{(D)}$ respectively.
%$G_{MN}$ is a block diagonal form as
%$G_{MN}={\rm diag}\{g_{\mu\nu}(x), \gamma_{mn}(y)\}$, $\phi=\phi_0$  and $N^\mu_m=0$, where $g_{\mu\nu}$ and $\gamma_{mn}$ are metrics defined in $d-$ and $D-$dimensions respectively and $x$ and $y$ are $d-$ and $D-$dimensional coordinate variables. $N^\mu_m=0$ is shift vector.
It is demonstrated that it can result in a huge size separation between the scales, e.g. the effective cosmological constants of $d$-dimensional spacetime and the extra $D$ dimensions, when $z=\frac{2-d}{D}+\epsilon$ and $\phi_0$ is of order one value, where $\epsilon$ is a small negative number and $\phi_0>1$(or  $\epsilon$ is a small positive number and $\phi_0<1$). 
For example, in the case of $d=4$ and $D=2$, the ratio of effective cosmological constants $\Lambda^{(d=4)}$ and $\Lambda^{(D=2)}$ which appears in the vacuum solution %defined in $4-$ and extra $1-$dimensions 
is given by
\begin{equation}
\Lambda^{(4)}\sim 10^{-124}\Lambda^{(2)},
\end{equation}
%and has no
%For example, when $z=\frac{2-d}{D}+\epsilon$, where $\epsilon$ is rather small and positive number, then $\phi_0^{4\frac{1-z}{d-2+zD}}=e^{ 4\frac{1-z}{d-2+zD}\log \phi_0}$
% becomes very much small by assuming that $\phi_0\sim 10$. If $\epsilon\sim-0.1$, then  
%\Lambda^{(d)}\sim 10^{-124}\Lambda^{(D)}$.}
when $\phi_0\sim 10$ and $\epsilon\sim-0.03$.
We note that if $\Lambda^{(2)}$ is the Planck scale, $\Lambda^{(4)}$ becomes near the current value of the cosmological constant in our universe.}
This type of argument can be used to generate  the vanishingly small cosmological constant \cite{Moon:2017rox}
and  photon mass \cite{Kim:2020ubm, Kouwn:2015cdw}.
%   Since anisotropic factor $z$ enters in the action as a free parameter, the isotropic case can be readily recovered by setting $z=1$. It seems that  the anisotropic   approach provides a road  to  a single framework to describe aspects of both  low-dimensional and higher-dimensional gravity theories  which cannot be unraveled in the isotropic approach. 

The paper is organized as follows; In Sec.2, a  brief review of FPD geometry is given and gravity theory in $(d, D)$ spacetime is constructed. Then,   anisotropic conformal extension is performed with  the use of the Weyl scalar field.
In Sec.\ref{Scale Emergence and Large Hierarchy}, scale emergence by conformal symmetry breaking and large hierarchy problem is discussed.
Suppression of the terms pertaining to extra dimensions are also given.
In Sec.\ref{Lower Dimensional Gravity}, aspects of lower-dimensional gravity is given and possible application to quantum gravity in four dimensions is sketched. In Sec.\ref{Vacuum Solution}, 
vacuum solutions are obtained and various cases of extra dimensions are discussed. 
Sec.6 contains summary and discussions.

%\vspace{-3cm}
\section{Anisotropic Conformal Gravity}
In this section, we start from a  comprehensive  review of FPD geometry  and anisotropic conformal gravity theory in $(d, D)$ spacetime is constructed. The presentation is an extended version of what is given in \cite{Moon:2017rox} and  is intended to make the paper self-contained. 
\subsection{ FPD  Invariant $(d, D)$ Gravity }
%\textcolor{blue}
We start with a brief review of $(d+D)$-dimensional gravity which respects the FPD invariance   \cite{Cho:1991wn, Yoon:2000sq}. 
The $(d+D)$-dimensional metric is expressed by employing ADM decomposition as
\begin{eqnarray}
ds^2&=&G_{AB}dX^A dX^B~~%(X^A= (x^\mu, y^m); A, B, ...=1,..., d+D)  
\\
&=&g_{\mu\nu}(dx^\mu +N^\mu_m dy^m)(dx^\nu +N^\nu_n dy^n)+\gamma_{mn}dy^m dy^n,\label{lineelement1} \nn
%&&(\mu, \nu, \cdots=0,1,\cdots, d;m,n,\cdots =d+1,\cdots, d+ D),\nn
\end{eqnarray}
where the capital Roman indices as $A$, $B$... run over $(d+D)$-dimensional spacetime, i.e. $ A, B, ...=1,..., d+D$. We use $X^A$ for the coordinate variable of the $(d+D)$-dimensional spacetime. The $(d+D)$-dimensional coordinate is a direct sum of $d$-dimensional spacetime and $D$-dimensional space coordinates as $X^A= (x^\mu, y^m)$, where the sub-indices $\mu$, $\nu$ run from $1$ to $d$ and $m$, $n$ run from $d+1$ to $d+D$. $G_{AB}$, $g_{\mu\nu}$ and $\gamma_{mn}$ are the metrics of $(d+D)$-, $d$- and $D$- dimensions respectively.
%coordinate $X^A$ is 
$N_m^{\mu}$ is the shift vector along the $d$-dimensional base spacetime corresponding to the $m$-th direction \cite{Misner:1974qy}. 

The inverse metric is obtained as 
\begin{eqnarray}
G^{AB}=\begin{pmatrix}g^{\mu\nu}+g^{mn}N_m^\mu N_n^\nu& -N^{\mu n} \\
-N^{m\nu} & \gamma^{mn} \\
\end{pmatrix},
\end{eqnarray}
where $g^{\mu\nu}$ and $\gamma^{mn}$ are inverse metrics of $g_{\mu\nu}$ and $\gamma_{mn}$ respectively. $N^{\mu m}=\gamma^{mn}N_n^\mu.$ 

Now, let us consider FPD and its coordinate transform being given by
\begin{eqnarray}\label{xytr}
x^{\mu}\to x^{\prime\mu}\equiv x^{\prime\mu}(x,y),~~~y^n\to y^{\prime n}\equiv
y^{\prime n}(y),\label{fFPD}
\end{eqnarray}
whose infinitesimal transformations can be written as \begin{eqnarray}
x^{\prime \mu}=x^\mu+\xi^\mu(x,y),~~~y^{\prime m}=y^{m}+\eta^m(y);~~~\partial_{\mu}
\eta^m=0,
\end{eqnarray}
where $\eta^m(y)$ is a function of $y^m$ only. 
%One can  check that
%\begin{eqnarray}
%\delta g_{\mu\nu}&=&-{\cal L}_{\xi}(x)g_{\mu\nu}-\eta^m(y)\partial_{m}g_{\mu\nu}
%\label{inft1},\\
%\delta N^{\mu}_{m}&=&-{\cal L}_{\xi}(x)N^{\mu}_{m}-
%{\cal L}_{\eta}(y)N^{\mu}_{m}-\partial_m\xi^{\mu},\label{inft2}\\
%\delta \gamma_{mn}&=&-{\cal
%L}_{\eta}(y)\gamma_{mn}-\xi^\mu\partial_{\mu}\gamma_{mn},\label{inft3}
%\end{eqnarray}
%where ${\cal L}_{\xi}(x)$ and ${\cal L}_{\eta}(y)$ with
%$\xi=\xi^{\mu}\partial_{\mu},~\eta=\eta^{m}\partial_{m}$ are the
%Lie-derivative acting only on the indices of $x$ and $y$,
%respectively.
%From the transformations (\ref{inft1})-(\ref{inft3}), 
The finite coordinate
transformations of each component of the metric under FPD\ of (\ref{fFPD})
are given as follows \cite{Moon:2017rox};
\begin{eqnarray}
g^{'}_{\mu\nu}(x',y')&=&\frac{\partial x^\rho}{\partial
x^{'\mu}}\frac{\partial
x^\sigma}{\partial x^{'\nu}}g_{\rho\sigma}(x,y),\label{trans1}\\
{N'}_{m}^{\mu}(x',y')&=&\Big(\frac{\partial y^n}{\partial
y^{'m}}\Big)\Big[\frac{\partial x^{'\mu}}{\partial
x^{\nu}}N_{n}^{\nu}(x,y)-\frac{\partial x'^{\mu}}{\partial
y^{n}}\Big],\label{trans2}\\
\gamma'_{mn}(x',y')&=&\frac{\partial y^p}{\partial
y^{'m}}\frac{\partial y^q}{\partial y^{'n}}\gamma_{pq}(x,y).\label{trans3}
\end{eqnarray}
One can easily check  that the above transformations (\ref{trans1})-(\ref{trans3}) leave the line element (\ref{lineelement1}) invariant.
%The covariant quantities (\ref{inva1})-(\ref{inva2})
%with the coordinates Introducing the horizontal vector
%fields $\hat{\partial}_m=\partial_m-N_{m}^{\mu}\partial_{\mu}$,
%can be contracted to generate FPD invariant  quantities.\begin{equation}
%\hat\partial_m= \partial_m -N_m^\mu\partial_\mu.
%\end{equation}

 %$\bullet$ more references on higher dimensional FPD invariant theory and %Kaluza-Klein in general(cho-freund).

For further computation, it is more convenient to introduce orthogonal basis in which the computation of the curvatures scalar $R$ is simplified to a great extent. We employ the orthogonal basis $\hat X^A = ( \hat x^\mu, \hat y^m)$\cite{Cho:1975sf} and  rewrite
line element (\ref{lineelement1}) as 
\begin{eqnarray}
ds^2= %\textcolor{blue}
{\hat G_{AB}d\hat X^A d \hat X^B}
\equiv\hat g_{\mu\nu}d\hat x^\mu d\hat x^\nu+\hat\gamma_{mn}d\hat y^md\hat y^n,\label{metricq}
\end{eqnarray}
where the transformations between the coordinate and the orthogonal bases are given by
\begin{equation}
\label{orthogonal-basis}
d\hat x^\mu=dx^\mu+N_m^\mu dy^m, ~~ d\hat y^m=dy^n.
\end{equation}
The relations between the metrics are
\begin{eqnarray}
\hat g_{\mu\nu}=g_{\mu\nu}, ~~ \hat \gamma_{mn}=\gamma_{mn}.
\end{eqnarray}
We note that these transformations are non-integrable. 
%%%%%%%%%%%2021.7.5 3시 30분%%%%%%%%%%%%%%%%%%%%%55

Next, we develop an algebra between translation operators(partial derivative operators) in orthogonal basis. %Going from orthogonal basis to coordinate basis can be formally done using
An observation of (\ref{orthogonal-basis}) suggests that
\begin{eqnarray}
\frac{\partial \hat x^\nu}{\partial x^\mu}=\delta^\nu_\mu,
~ \frac{\partial \hat y^m}{\partial x^\mu}=0,
~\frac{\partial \hat x^\nu}{\partial y^m}=N^\nu_m,
~\frac{\partial \hat y^m}{\partial y^n}=\delta^m_n,~
\end{eqnarray}
and their inverse relations, 
\begin{eqnarray}
\frac{\partial  x^\nu}{\partial \hat x^\mu}=\delta^\nu_\mu,
~ \frac{\partial  y^m}{\partial \hat  x^\mu}=0,
~\frac{\partial  x^\nu}{\partial \hat y^m}=-N^\nu_m,
~\frac{\partial  y^m}{\partial \hat y^n}=\delta^m_n.
\end{eqnarray}
By using these facts, the derivative operators in orthogonal basis is given by
\begin{eqnarray}
\frac{\partial}{\partial\hat x^\mu}=\frac{\partial}{\partial x^\mu},
~~\frac{\partial}{\partial\hat y^m}=\frac{\partial}{\partial y^m}-
N^\mu_m\frac{\partial}{\partial x^\mu},
\end{eqnarray}
in terms of the old variables.
We also have formulas for their second derivatives as
\begin{eqnarray}
\frac{\partial}{\partial\hat x^\mu}\frac{\partial}{\partial\hat x^\nu}
=\frac{\partial}{\partial\hat x^\nu}\frac{\partial}
{\partial \hat x^\mu},~~\frac{\partial}{\partial\hat x^\mu}
\frac{\partial}
{\partial \hat y^m}-\frac{\partial}{\partial\hat y^m}
\frac{\partial}{\partial \hat x^\mu}=
-\frac{\partial N^\nu_m}{\partial\hat x^\mu}
\frac{\partial}{\partial \hat x^\nu}
\end{eqnarray}
and
\begin{equation}
\frac{\partial}{\partial\hat y^m}\frac{\partial}{\partial\hat y^n}-\frac{\partial}{\partial\hat y^n}\frac{\partial}{\partial\hat y^m}=-F^\mu_{mn}\frac{\partial}{\partial\hat x^\mu},
\end{equation}
where
\begin{eqnarray}
F^{\mu}_{mn} &=&\partial_{m}N_{n}^{\mu}-\partial_{n}N_{m}^{\mu}-N_{m}^{\nu}\partial_{\nu}N_{n}^{\mu}
+N_{n}^{\nu}\partial_{\nu}N_{m}^{\mu}.\label{fdef}
\end{eqnarray}
We package these relations as a compact form of larger commutation relations as
%which defines the structure functions
\begin{eqnarray}
[\hat\partial_A, \hat\partial_B]=f^C_{~AB}\hat\partial_C ~~,
\end{eqnarray}
where the structure constant $f^C_{~AB}$ is given by
\begin{eqnarray}
f^m_{~\mu\nu}=f^\sigma_{~\mu\nu}=0, 
~f^\nu_{~\mu m}=-f^\nu_{~m\mu}=
-\frac{\partial N^\nu_m}{\partial \hat x^\mu},~f_{~mn}^\mu=-F_{mn}^\mu.
\end{eqnarray}
Geometrically, the field strength $F^{\mu}_{mn}$ is a Diff(d) vector-valued curvature associated with the holonomy of the shift vector $N_n^\mu$ along the extra dimensions.

%%%%%%%%%%%2021.7.5 6시 10분%%%%%%%%%%%%%%%%%%%%%55

Now we are ready to compute curvature tensor in $d+D$ dimension. First, we define the connections. To do this, we introduce vectors
% $\vec r =(x_0, x_1, \cdots , x_{d+D})$ $\rightarrow\textcolor{blue}
${\vec r =(X^0, X^1, \cdots , X^{d+D})}$
%for convenience 
and %consider a vector 
$\vec V$ in the basis 
%$\vec {\hat e}_A\equiv \frac{\partial \vec r}{\partial \hat x^A}$, 
%$\rightarrow \textcolor{blue}
${\vec {\hat e}_A\equiv \frac{\partial \vec r}{\partial \hat X^A}}$ such that 
$\vec V=\hat V^A \vec {\hat e}_A$. These basis vectors are directly related with the metric coefficients of  (\ref{metricq}) by
\begin{equation}
 \vec {\hat e}_A\cdot \vec{\hat e}_B=\hat G_{AB}.\label{metricqq}
\end{equation}
Equation for parallel transport is defined as    
\begin{eqnarray}
\frac{d\vec V}{du}=\frac{d\hat X^A}{du}\left[
\frac{\partial \hat V^B}
{\partial \hat X^A}\vec {\hat e}_B+
\hat V^B\frac{\partial\vec{\hat e}_B}{\partial \hat X^A}\right]=0
\end{eqnarray}
%%%%%%%%%%%2021.7.6 10시 11분%%%%%%%%%%%%%%%%%%%%%55
along a curve parametrized by an affine parameter $u.$
The next step is to introduce connections in this orthogonal coordinate by
\begin{eqnarray}
\frac{\partial \vec{\hat e}_B}{\partial \hat X^A}=\hat\Gamma^C_{~AB}\vec{\hat e}_C,
\end{eqnarray}
and using (\ref{metricqq}), we obtain %and using $\vec {\hat e}_A\cdot \vec{\hat e}_B=\hat g_{AB}$, one finds 
\begin{eqnarray}
\hat\Gamma^C_{~AB}=\hat\Gamma^{C}_{(C)AB}+
\frac{1}{2}\hat G^{CD}\Big(f_{DAB}-f_{ABD}-f_{BAD}\Big), ~~(f_{ABC}=\hat G_{AD}f^D_{~BC}),
\end{eqnarray}
where $\hat\Gamma^{C}_{(C)AB}$ is the usual form of Chrisoffel connection being given by
\begin{equation}
%\textcolor{blue}
{\hat\Gamma^{C}_{(C)AB}=\frac{1}{2}\hat G^{CD}\left(\frac{\partial \hat G_{DA}}{\partial \hat X^B}+ \frac{\partial \hat G_{DB}}{\partial \hat X^A}- \frac{\partial \hat G_{AB}}{\partial\hat X^D}\right)}.
\end{equation}
The covariant derivative in this orthogonal coordinate is defined with the connection $\hat\Gamma^C_{~AB}$ by
\begin{eqnarray}
\hat\nabla_A \hat V^B\equiv\frac{\partial \hat V^B}
{\partial \hat X^A}+\hat\Gamma^B_{~AC}\hat V^C,
\end{eqnarray}
and one can prove explicitly the metric compatibility,
\begin{eqnarray}
\hat\nabla_A\hat G_{BC}=0.
\end{eqnarray}

The Riemann tensor is defined as 
\begin{eqnarray}
[\hat\nabla_A, \hat\nabla_B]\hat V^C=\hat R^C_{~DAB}\hat V^D,
\end{eqnarray}
where
\begin{eqnarray}
\hat R^A_{~BCD}=\hat\partial_C\hat\Gamma^A_{~DB}
-\hat\partial_D\hat\Gamma^A_{~CB}+
\hat\Gamma^A_{~CE}\hat\Gamma^E_{~DB}
-\hat\Gamma^A_{~DE}\hat\Gamma^E_{~CB}-f^E_{~CD}\hat\Gamma^A_{~EB}.
\end{eqnarray}
A straightforward computation  yields the scalar curvature \cite{Yoon:1996uu} (up to  total derivative terms)
 \begin{eqnarray}
\label{CRuv-invv}
 R=\hat G^{BD}\hat R^A_{~BAD}
 =R^{(d)}+\hat R^{(D)}+R_F+R_K+R_R, \label{fiveR}
 \end{eqnarray}
 where  $R^{(d)}=g^{\mu\nu}R_{\mu\nu}$ is the usual scalar curvature constructed with Christoffel connection
$\Gamma^\mu_{\nu\rho}(g_{\alpha\beta})=\frac{1}{2}g^{\mu \eta}\left(\frac{\partial g_{\eta\nu}}{\partial x^\rho}+ \frac{\partial g_{\eta\rho}}{\partial x^\nu}- \frac{\partial g_{\mu\nu}}{\partial x^\eta}\right)$, 
being given by
\begin{eqnarray}
{ R}_{\mu\nu}=\partial_{\alpha}{\Gamma}_{\mu\nu}^{\alpha}- \partial_{\nu}{\Gamma}_{\alpha \mu}^{\alpha}+{\Gamma}_{\alpha \beta}^{\beta}{\Gamma}_{\mu\nu}^{\alpha}
-{\Gamma}_{\alpha \nu}^{\beta}{\Gamma}_{\beta \mu}^{\alpha}.
\end{eqnarray}
%and \textcolor{blue}
{$\hat R^{(D)}=\gamma^{mn}R_{mn}$, where
%constructed with 
%with $\Gamma^m_{np}$ with the $\gamma_{mn}$ and 
%the derivative 
\begin{eqnarray}
\hat{ R}_{mn}=\hat\partial_{p}\hat{\Gamma}_{nm}^{p}-\hat \partial_{n}\hat{\Gamma}_{pm}^{p}+\hat{\Gamma}_{pq}^{q}\hat{\Gamma}_{nm}^{p}
-\hat{\Gamma}_{pn}^{q}\hat{\Gamma}_{qm}^{p},
\end{eqnarray}
with the connection $\hat{\Gamma}_{mn}^{p}$ which is given by
\begin{eqnarray}
\hat{\Gamma}_{mn}^{p}=\frac{1}{2}\gamma^{pq}(\hat{\partial}_m\gamma_{nq}
+\hat{\partial}_n\gamma_{qm}-\hat{\partial}_q\gamma_{mn}).
\end{eqnarray}}
%\nointend
Again, the derivative operator, $\hat \partial_m\equiv \frac{\partial}{\partial \hat y^m}
=\partial_m-N^\mu_m\partial_\mu$.
The curvature scalars, $R^{(d)}=g^{\mu\nu}R_{\mu\nu}$ and
$\hat{R}^{(D)}=\gamma^{mn}\hat{R}_{mn}$ are
FPD invariants. To show these, one may use a fact, 
\begin{equation}
\hat{\partial}^{'}_m=\left(\frac{\partial y^n}{\partial y^{'m}}\right)
\hat\partial_n.
\end{equation}
In addition to these two curvature scalars, $R^{(d)}$ and $\hat R^{(D)}$,  we also have three more FPD invariants,
\begin{eqnarray}
R_F&=&-\frac{1}{4}\gamma^{mn}\gamma^{pq}g_{\mu\nu}F_{mp}^{\mu}F_{nq}^{\nu},
\nn\\
R_K&=& -\frac{1}{4}\gamma^{mn}g^{\mu\nu}g^{\alpha\beta}({\cal
D}_m g_{\mu\alpha}{\cal D}_{n}g_{\nu\beta}
 -{\cal D}_m g_{\mu\nu}{\cal D}_{n}g_{\alpha\beta}),\nonumber\\
R_R &=&-\frac{1}{4}g^{\mu\nu}\gamma^{mn}\gamma^{pq}(\partial_{\mu}
\gamma_{mp}\partial_{\nu}\gamma_{nq}
 -\partial_{\mu}
\gamma_{mn}\partial_{\nu}\gamma_{pq}),\label{each}
\end{eqnarray}
where the symbol, ${\cal D}_{n}g_{\mu\nu}$ is the Diff(d)-covariant derivative  defined by \cite{Cho:1991wn, Yoon:2000sq}
% \footnote{ FPD in Ref. \cite{Cho:1991wn, Yoon:2000sq} is restricted to %the case where $y^{\prime n}=
%y^{n}, \eta^n=0$}:
\begin{eqnarray}\label{ddef}
{\cal D}_{m}g_{\mu\nu}&=&\partial_{m}g_{\mu\nu}-N^{\rho}_{m}\partial_{\rho}g_{\mu\nu}-
(\partial_{\mu}N^{\rho}_{m})g_{\rho\nu}
-(\partial_{\nu}N^{\rho}_{m})g_{\mu\rho}\nn.
%&=&
%\nabla_m g_{\mu\nu}-\nabla_\mu N_{m\nu}-\nabla_\nu N_{m\mu},\\
%{\cal F}_{mn}&=&[{\cal D}_m,{\cal D}_{n}],~~(\equiv {\cal F}^{\mu}_{mn}\partial_\mu) %\nn\\
\end{eqnarray}
A geometrical interpretation for the ${\cal D}_m g_{\mu\nu}$ is that it is a kind of generalized extrinsic curvature along the $m$-th extra direction. 
One can verify that under FPD transform given in (\ref{xytr}),  ${\cal D}_m g_{\mu\nu}$ and
$F_{mn}^{\mu}$  change covariantly as
\begin{eqnarray}\label{dftr}
({\cal D}_m g_{\mu\nu})^{'}&=&\frac{\partial y^{n}}{\partial
y^{'m}}\frac{\partial x^{\rho}}{\partial x^{'\mu}}\frac{\partial
x^{\sigma}}{\partial x^{'\nu}}{\cal D}_n g_{\rho\sigma},\label{inva1}\\
(F^{\mu}_{mn})^{'}&=&\frac{\partial y^{p}}{\partial
y^{'m}}\frac{\partial y^{q}}{\partial y^{'n}}\frac{\partial
x^{'\mu}}{\partial x^{\nu}}F^{\nu}_{pq}.\label{inva2}
\end{eqnarray}
%
%
%
%
%%%%%%%%%%%2021.7.7 8시 34분%%%%%%%%%%%%%%%%%%%%%
%
%%%%%%%%%%%%2021.7.10 7시 45분%%%%%%%%%%%%%%%%%%%%%
%
%One also has the spacetime derivative of the metric $\gamma_{mn}$ transforming as
The transformation of $\gamma_{mn}$ is also covariant, i.e.
\begin{eqnarray}
(\partial_\mu \gamma_{mn})^{'}&=&\frac{\partial x^{\nu}}{\partial
x^{'\mu}}\frac{\partial y^{p}}{\partial y^{'m}}\frac{\partial
y^{q}}{\partial y^{'n}}(\partial_\nu \gamma_{pq})\label{inva3}
\end{eqnarray}
under FPD transformation, where to derive that, one may use a fact, ${\partial}^{'}_{\mu}=\left(\frac{\partial x^{\nu}}{\partial x^{'\mu}}\right)
\partial_{\nu}.$
%of (\ref{xytr}), because ${\partial}^{'}_{\mu}=\left(\frac{\partial x^{\nu}}{\partial x^{'\mu}}\right) \partial_{\nu}.$
Therefore, we conclude that each of the five terms in (\ref{fiveR})
is separately FPD invariant. 
%
%Also, using 
%\begin{equation}
%\hat{\partial}^{'}_m=\left(\frac{\partial y^n}{\partial y^{'m}}\right)
%\hat\partial_n
%\end{equation}
%under FPD,
%we find that both the curvature
%scalar  $R^{(4)}=g^{\mu\nu}R_{\mu\nu}$ and
%$\hat{R}^{(D)}=\gamma^{mn}\hat{R}_{mn}$ are
%FPD  invariants.  
%
%
%
%
%
%%%%%%%%%%%2021.7.7 9시 4분%%%%%%%%%%%%%%%%%%%%%
%
%
%

In the end, we can construct a general FPD invariant $(d+D)$-dimensional action by employing a linear combination of the five FPD invariants given in (\ref{CRuv-invv}). In fact, the number of terms in the linear combination becomes eight since cosmological constant $\Lambda$ is manifestly FPD invariant as well as  the two terms in $R_K$ and the other two in $R_R$ are separately FPD invariant:
\begin{eqnarray}\label{conacd}
S^{(d+D)}&=&M_*^{d-2+D}\int d^{4}xd^Dy \sqrt{-g}\sqrt{\gamma}\Big[
\left( R^{(d)}%\textcolor{
%-\textcolor{2\Lambda}
-2\Lambda
%}
\right)+\alpha_{D}\hat{R}^{(D)}
\nn\\
&&-\frac{\alpha_F}{4}\gamma^{mn}\gamma^{pq}g_{\mu\nu}F_{mp}^{\mu}F_{nq}^{\nu}
 -\frac{\alpha_K}{4}\gamma^{mn}g^{\mu\nu}g^{\alpha\beta}({\cal
D}_m g_{\mu\alpha}{\cal D}_{n}g_{\nu\beta}
 -\alpha{\cal D}_m g_{\mu\nu}{\cal D}_{n}g_{\alpha\beta})\nonumber\\
 &&-\frac{\alpha_R}{4}g^{\mu\nu}\gamma^{mn}\gamma^{pq}(\partial_{\mu}
\gamma_{mp}\partial_{\nu}\gamma_{nq}
 -\beta\partial_{\mu}
\gamma_{mn}\partial_{\nu}\gamma_{pq})\Big] +S_{m},\label{ehq}
\end{eqnarray}
where  we fix the coefficient of $R^{(d)}$ to be one and $\alpha_D,\alpha_F,\alpha_K,\alpha_R,\alpha,$ and $\beta$ are arbitrary constants.
$M_*^{d-2+D}$ is the higher-dimensional gravitational constant. $S_m$ is the matter action which will not be considered in this work.
One can show that when all of $\alpha_D,\alpha_F,\alpha_K,\alpha_R,\alpha,$ and $\beta$ become unity, all the terms in the action (\ref{ehq})
combine into $(d+D)$-dimensional Einstein-Hilbert action with cosmological constant where FPD get elevated to the full $(d+D)$-dimensional  general diffeomorphism.

%\textcolor{blue}
{Especially, we note that
when $d=4$ and $D=1$, %we have $\alpha_D=\alpha_F=0$ and $m=n=5.$
$\hat R^{(D)}$ and $F^\mu_{mn}$ vanish and $m=n=5$.
With a definition of $\gamma_{55}=N^{2},$ the action (\ref{ehq}) becomes $(4+1)$-dimensional 
%FPD-invariant anisotropic gravity \footnote {Compare with the action given in 
%\cite{Papantonopoulos:2011xa}. See also Ref. \cite{bemfica} for time-anisotropic $(4+1)$-dimensional theory.    }
%\begin{eqnarray}
%&&\hspace*{-2em}S^{(5)}=M_{*}^{3}\int 
%d^4x dy\sqrt{-g}N\Bigg[\left(R^{(4)}-2\Lambda
%\right)
%-\alpha_K\Big(K_{\mu\nu}K^{\mu\nu} -
%\alpha K^2\Big)+\alpha_R(1-\beta) \frac{\nabla_{\mu}N\nabla^{\mu}N}{N^2}\Bigg],\nn \\
%{\ }
%\end{eqnarray}
%where $K_{\mu\nu}$ is the extrinsic curvature tensor,
%\begin{equation} 
%K_{\mu\nu}=(\partial_y g_{\mu\nu} - \nabla_{\mu} N_{\nu} -
%\nabla_{\nu} N_{\mu})/(2N).
%\end{equation}
FPD-invariant anisotropic gravity
\begin{eqnarray}
\hspace*{-2em}S^{(5)}&=&M_{*}^{3}\int 
d^4x dy\sqrt{-g}N\Bigg[\left(R^{(4)}-2\Lambda
\right)
-\alpha_K\Big(K_{\mu\nu}K^{\mu\nu} -
\alpha K^2\Big)\nn\\&&~~~~~~~~~~~~~~~~~~~~~~~~~~~~~~~~
+\alpha_R(1-\beta) \frac{\nabla_{\mu}N\nabla^{\mu}N}{N^2}\Bigg],\label{foursome}
\end{eqnarray}
where $K_{\mu\nu}$ is the extrinsic curvature tensor,
\begin{equation} 
K_{\mu\nu}=(\partial_y g_{\mu\nu} - \nabla_{\mu} N_{\nu} -
\nabla_{\nu} N_{\mu})/(2N).
\end{equation}
We note that the action (\ref{foursome}) is 4+1 dimensional Horava gravity theory but it is a little more general form\footnote{
Higher-dimensional Horava-type gravity theories have been studied extensively. For example, there are theories adapting foliation along the time
dimension\cite{Horava:2009if}, studies of black hole solutions \cite{Koutsoumbas:2010pt,Lin:2014nua}  and with the braneworld models \cite{bemfica} in this framework.  
}. % up to   quadratic kinetic terms where the extra spatial dimension has
%enjoying anisotropic scaling symmetry.
%than the spacetime four dimensions. 
For example, one can compare this action with equation (2.3) in \cite{Papantonopoulos:2011xa}, which is recovered from (\ref{foursome}) when we set $\beta=1$.
%\footnote{However, compare the action (\ref{foursome}) with (2.3) given in \cite{Papantonopoulos:2011xa}. }. 
% and $\alpha_7 =\alpha_5 (1-\alpha_6)$.}

% $S_m$ is the matter action whose constituents will not be our concern.

%, but we will compute general constraint on the energy-momentum tensor of %$S_m$ imposed by FPD invariance in 5D case.
% For generic values of $\alpha$'s  the theory exhibits pathological behaviors %displaying ghost-like excitations in the perturbation theory \cite{Papantonopoulos:2011xa}. %This issue will be taken up later  for the  5D conformally invariant case.
%%%%%%%%%%%%%%%%%%%%%%%%%%%%%%%%%%%%%%%%%%%%%%%%%%%%%%%%%%%%%%
%%%%%%%%%%%%%%%%%%%%%%%%%%%%%%%%%%%%%%%%%%%%%%%%%%%%%%%%%%%%%%
%%%%%%%%%%%%%%%%%%%%%%%%%%%%%%%%%%%%%%%%%%%%%%%%%%%%%%%%%%%%%%
%
%
%
%
%
%%%%%%%%%%%2021.7.10 9시 51분%%%%%%%%%%%%%%%%%%%%%
%
%
%
\subsection{Conformal extension}
\label{Conformal extension}
%%%%%%%%%%%%%%%%%%%%%%%%%%%%%%%%%%%%%%%%%%%%%%%%%%%%%%%%%%%%%%
%%%%%%%%%%%%%%%%%%%%%%%%%%%%%%%%%%%%%%%%%%%%%%%%%%%%%%%%%%%%%%
%%%%%%%%%%%%%%%%%%%%%%%%%%%%%%%%%%%%%%%%%%%%%%%%%%%%%%%%%%%%%%

In this subsection, we consider a conformal extension of the action(\ref{conacd}) which can be achieved  by introducing an additional scalar field(Weyl scalar field) $\phi$ into it \cite{Moon:2009zq}. To assign  anisotropic scaling symmetry between the $d$-dimensional spacetime and extra $D$ dimensions,
one needs to introduce an anisotropic scaling parameter ``$z$" which will characterize the anisotropy. 
%of the extra dimensions and 
Let us consider the following conformal transformation
\begin{eqnarray}\label{weyltf}
g_{\mu\nu}\rightarrow e^{2\omega(x,y)}g_{\mu\nu},~~
\gamma_{mn}\rightarrow e^{2z\omega(x,y)}\gamma_{mn},~~ \phi\rightarrow
e^{-v(z)\omega(x,y)}\phi,~~N_{m}^{\mu}\rightarrow N_{m}^{\mu}.\label{confotrans1}
\end{eqnarray}
Here, $v(z)$ can be an arbitrary function of $z$ which can be adjusted with a suitable redefinition of the field $\phi$. We take $v=\frac{d-2+zD}{2}$ in order to set the conformal factor in front of $R^{(d)}$ to be $\phi^2$.
In other wards, the anisotropic scaling weights of the fields given in (\ref{confotrans1}) are
\begin{eqnarray}
[g_{\mu\nu}]=2, ~~[\gamma_{mn}]=2z, ~~[\phi]=-\frac{d-2+zD}{2}, 
~~[N^\mu_m]=0.
\end{eqnarray}
$z=1$ corresponds to the isotropic conformal transformations
\cite{Moon:2009zq}.

Taking into account of the additional scalar field $\phi$, one can construct the anisotropic scale invariant action in $d+D$ dimensions by considering scale invariant combinations
%\begin{eqnarray}
%\tilde g_{\mu\nu}=\phi^{\frac{4}{d-2+zD}}g_{\mu\nu}, 
%~\tilde\gamma_{mn}=\phi^{\frac{4z}{d-2+zD}}\gamma_{mn},
%\end{eqnarray}
\begin{eqnarray}
\phi^{\frac{4}{d-2+zD}}g_{\mu\nu}, {\ \ \rm\ and\ \ }
\phi^{\frac{4z}{d-2+zD}}\gamma_{mn}.
\end{eqnarray}
In the FPD invariant action (\ref{conacd}), we replace every $g_{\mu\nu}$, and $\gamma_{mn}$ by $\phi^{\frac{4}{d-2+zD}}g_{\mu\nu}$ and $\phi^{\frac{4z}{d-2+zD}}\gamma_{mn}$ respectively, then one obtains
\begin{eqnarray}
&&S=\int d^{d}xd^{D}y \sqrt{-g}\sqrt{\gamma}
\Bigg[{\cal L}_d+{\cal L}_D+{\cal L}_F+{\cal L}_K+{\cal L}_R\Bigg],\label{taction}\\
&&{\cal L}_{d}=
 \phi^2\Big(R^{(d)}-4\frac{d-1}{d-2+zD}\frac{\nabla_{\mu}\nabla^{\mu}\phi}{\phi}
+4\frac{zD(d-1)}{(d-2+zD)^2}\frac{\nabla_{\mu}\phi\nabla^{\mu}\phi}{\phi^2}
\Big)-V_0\phi^{2\frac{d+zD}{d-2+zD}},\nn\\
 &&{\cal L}_D=\alpha_D\phi^{2\frac{d+z(D-2)}{d-2+zD}}
 \Big(\hat{R}^{(D)}-4z
\frac{D-1}{d-2+zD}\frac{\hat{\nabla}_{m}\hat{\nabla}^{m}\phi}{\phi}+
4\frac{z(D-1)(d-2+2z)}{(d-2+zD)^2}\frac{\hat{\nabla}_{m}\phi\hat{\nabla}^{m}\phi}{\phi^2}\Big),
\nn\\
&&{\cal L }_F= -\frac{\alpha_F}{4}\phi^{2\frac{d+2+z(D-4)}{d-2+zD}}
\gamma^{mn}\gamma^{pq}g_{\mu\nu}F_{mp}^{\mu}F_{nq}^{\nu},\nn\\
 &&{\cal L}_K=-\frac{\alpha_K}{4}\phi^{2\frac{d+z(D-2)}{d-2+zD}}\gamma^{mn}g^{\mu\nu}g^{\alpha\beta}
 \left(\tilde{\cal
D}_m g_{\mu\alpha}\tilde{\cal D}_{n}g_{\nu\beta}
 -\alpha\tilde{\cal D}_m g_{\mu\nu}
 \tilde{\cal D}_{n}g_{\alpha\beta}\right),\nn\\
&&{\cal L}_R=-\frac{\alpha_R}{4}\phi^{2}
g^{\mu\nu}\gamma^{mn}\gamma^{pq}\left(
\tilde\nabla_{\mu}
\gamma_{mp}\tilde\nabla_{\nu}\gamma_{nq}
 -\beta\tilde\nabla_{\mu}
\gamma_{mn}\tilde\nabla_{\nu}\gamma_{pq}\right).\nn
\label{confact1}
\end{eqnarray}
In the above, $V_0=2\Lambda$ is a constant and the hatted covariant derivative is given by
$\hat{\nabla}_m j^m=\hat{\partial}_m j^m+\hat{\Gamma}_{mp}^m
j^p$ and 
\begin{eqnarray}
\tilde{\cal D}_m g_{\mu\nu}=\left({\cal D}_m +
\frac{4}{d-2+zD}\frac{\hat\nabla_m\phi}{\phi}\right)g_{\mu\nu},
 \tilde\nabla_{\mu} \gamma_{mn}=
\left(\nabla_\mu +
\frac{4z}{d-2+zD}\frac{\nabla_\mu\phi}{\phi}\right)\gamma_{mn}.\nn
\end{eqnarray}
$\alpha_D, \alpha_F, \alpha_K, \alpha_R, \alpha$ and $\beta $ are also constants.
%\textcolor{blue}{ but they could in principle have $z$-dependence with normalization such that they all become unity at $z=1.$  }
The action (\ref{taction}) is the $(d+D)$-dimensional anisotropic Weyl gravity \cite{Moon:2009zq}.
% Note that the invariance under (\ref{confotrans1}) is built in the above action (\ref{taction}).
In the action (\ref{taction}), there are no dimensionful parameters at all. 

We note that this action also enjoys the following scaling symmetry:
\begin{eqnarray}
x\rightarrow b^{-1}x, ~~y\rightarrow b^{-z}y,
~~\phi\rightarrow 
b^{\frac{d-2+zD}{2}}\phi,~~N^\mu_m
\rightarrow b^{z-1}N^\mu_m.\label{st11}
\end{eqnarray}
%This gives $z$-dependent scaling dimensions  as 
The scaling dimensions of the  objects in the action are given by
\begin{eqnarray}
[x]_s=-1, ~~[y]_s=-z, ~~[\phi]_s=\frac{d-2+zD}{2}, ~~[N^\mu_n]_s=z-1,
\end{eqnarray}
and  all others have vanishing scaling dimensions.
% On the other hand, the anisotropic conformal invariance is given by (\ref{confotrans1}) and conformal weights are
%\begin{eqnarray}
%[g_{\mu\nu}]=2, ~~[\gamma_{mn}]=2z, ~~[\phi]=-\frac{d-2+zD}{2}, 
%~~[N^\mu_m]=0.
%\end{eqnarray}
%
%
%
%
%
%%%%%%%%%%%%%%%%2021.7.11 10시 5분%%%%%%%%%%%%%%%%%%%%%
%
%
%
%
%
%
%
%%%%%%%%55
\section{Scale Emergence and Large Hierarchy}
\label{Scale Emergence and Large Hierarchy}
\subsection{Scale Emergence in $(d+D)$-dimensional Weyl Gravity}
In the previous section, we discuss how the scale invariant action can be obtained from the FPD invariant action. {In fact, there is a $(d+D)$-dimensional gravitational constant, $M_*^{d-2+D}$ as an overall coefficient in the FPD action(\ref{ehq}), but we remove this in its conformal extension (\ref{taction}). The mass dimension associated to the gravitational constant is distributed to the coordinate variable $y$ and the Weyl scalar field $\phi$. In other words, to recover the FPD action(\ref{ehq}) from the conformal Weyl gravity action(\ref{taction}), rescale the coordinate $y$ and the field $\phi$ as
\begin{equation}
y\rightarrow M_*^{-z+1}y, ~~\phi\rightarrow M_*^{\frac{d-2+zD}{2}}\phi,\label{res11}
\end{equation}
and gauge-fix $\phi$ to  $\phi_0=1$ by exploiting the conformal invariance (independent of the value of $z$!).
}
%We  assume that  the field $\phi$ and the coordinate $y$  recover their canonical dimensions with the vacuum solution, and therefore  re-scale them via
%\begin{equation}
%y\rightarrow M_*^{-z+1}y, ~~\phi\rightarrow M_*^{\frac{d-2+zD}{2}}\phi,
%\end{equation}
This scaling assigns inverse-mass dimension to the coordinate $y$ and makes $\phi$ dimensionless respectively.

\subsection{Large Hierarchy from broken scale symmetry in $(d+D)$-dimensional Weyl Gravity}
Now we are somewhere in the middle between the FPD gravity and its conformal extension. As explained, one can set the field $\phi$ have vacuum expectation value $\phi_0$ by employing an appropriate symmetry breaking mechanism. However, $\phi_0$ does not necessarily become unity. $\phi_0$ can be designed to have some other values. 

We also have the anisotropic scale parameter $z$ in our theory. %Interplay 
Somehow with a mild control of these two parameters, $\phi_0$ and $z$, our model can reach other interesting destinations. One of them is resolution of hierarchy problem.
%Our model also suggests a resolution to the hierarchy problem. 
Let us consider a scalar  matter field $\Phi$(for simplicity) under the FPD and the scale transformation (\ref{st11}) and it lives in the $d$-dimensional manifold, i.e. $\Phi\rightarrow\Phi(x^\mu)$. Assume that the matter field action is given by
\begin{equation}
S_{matter}=\int d^dxd^Dy\sqrt{-g}\sqrt{\gamma}\left[ \frac{\phi^2}{2}g^{\mu\nu}\partial_\mu \Phi \partial_\nu \Phi- \frac{M_*^2}{2}\phi^{\frac{2(d+Dz)}{d-2+Dz}}\Phi^2\right],
\end{equation}
where the natural mass scale of the matter field $\Phi$ is $M_*$ which is the only scale appearing in our model(The rigorous derivation of this action from $(d+D)$-dimensional matter fields action is given in the Appendix). This action is manifestly invariant under the diffeomorphism and the scale transformation.
%The field $\Phi$ is a scalar under the the foliation preserving defeomorphism and the scale transformation for the action to respect those symmetries. 
As we discuss above, let us assume that the field $\phi$ has a vev, $\phi_0$. To keep the kinetic part to be canonical, take a field redefinition of the $\Phi$ as $\Phi=(\sqrt{V_D}\phi_0)^{-1}\tilde\Phi$, where $V_D=\int \sqrt{\gamma}d^Dy$. 
By integrating the $D$-dimensional volume, one can construct $d$-dimensional effective action of the field $\tilde\Phi$.
After all, one realizes that the physical mass of the field $\tilde\Phi$ becomes
$M_*\exp\left({\frac{2}{d-2+Dz}\log\phi_0}\right)$. The factor, %$\phi_0^{\frac{2}{d-2+Dz}}=$
$\exp\left({\frac{2}{d-2+Dz}\log\phi_0}\right)$ can give a large scale distinction between $M_*$ and the physical mass of the field $\tilde \Phi$ for $z=\frac{2-d}{D}+ \epsilon$ where $\epsilon$ is a small(but not very much small) and positive number with order one value of $\phi_0<1$(or $\epsilon$ is negative number with order one value $\phi_0>1$).

Comparison with RS model\cite{Randall:1999vf,Randall:1999ee} will provide a clearer interpretation of our result.
RS model gives rise to a large hierarchy between two scales appearing in the theory. The geometric setting is pretty much different from ours. They consider a 5-dimensional space of which the topology is given by $M_4\times S^1$, where the space $M_4$ is a 4-dimensional Minkowskian manifold.
The circle $S^1$ is parameterized by an angle $\psi$ and they put extended objects, which are called branes at $\psi=0$ and $\psi=\pi$. The brane at $\psi=0$ is given a name as `visible brane', where we live in and the one at $\phi=\pi$ is called `hidden brane', which we cannot observe. The solution of the Einstein equation of  the system is "so called" $\it warp$ geometry which is characterized by the warp factor $e^{-2kr_c|\psi|}$. 
Here, $k=\sqrt{-\frac{\Lambda}{24M^3}}$ with $\Lambda$ being the 5-dimensional (negative) cosmological constant and  $M$ is the 5-dimensional energy scale which is related to the 5-dimensional Newton constant as $G_5\sim M^{-3}$.
It turns out that any physical mass scale $m_{phy}$ in the 4-dimensional effective theory on the visible brane is suppressed by the warp factor from the mass parameter, $m_0$ appearing in the visible sector theory as
 $m_{phy}=e^{-kr_c\pi}m_0$.

The resolution of the hierarchy problem in RS model is achieved by tuning the factor $k r_c$. If $k r_c=15/\pi\sim 4.8$, then we get electro-weak scale from $m_0$ by assuming that $m_0$ is the Planck scale. On the other hand, combination of the scaling parameter $z$ and vev $\phi_0$ play the same role with the factor $k r_c$ in RS model.
%the large hierarchy occurs in our model due to the scaling factor $z$, which is somewhat tuned. 
If $\frac{2\log \phi_0}{d-2+Dz}\sim -15$, then we get electro-weak scale from $M_*$ by assuming that $M_*$ is the Planck scale.

We stress that such a large scale separation is obtained by a mild control of our vev $\phi_0$ and parameter $z$, not by a specific fine tuning of those parameters. For example, to get $\frac{2\log \phi_0}{d-2+Dz}\sim -15$ in $d=4$ and $D=1$, one may suppose that $\phi_0\sim e=2.71828...$(Euler number), being an order one value, and $z\sim -2.133$ \cite{Kouwn:2017qet}.
 %Other case is $\phi_0\sim e^{-1}$ and $z\sim -1.867$

%%%%%%%%%%%2021.7.15 오후 3시 45분%%%%%%%%%%%%%%%%%%%%%

\subsection{Suppression of the Lagrangian density}
\label{Suppression of the Lagrangian density}
Huge suppression  can occur in the anisotropic Weyl gravity action(\ref{taction}), once the Weyl scalar has a vev, $\phi_0$. With the scalar $\phi $
being fixed to its  $\phi_0$, the Weyl gravity action becomes
\begin{eqnarray}
&&S[\phi_0]=\int d^{d}xd^{D}y \sqrt{-g}\sqrt{\gamma}
\Bigg[{\cal L}_d+{\cal L}_D+{\cal L}_F+{\cal L}_K+{\cal L}_R\Bigg],\label{taction2}\\
&&{\cal L}_{d}=
 \phi^2_0R^{(d)}
-V_0\phi^{2\frac{d+zD}{d-2+zD}},\ \ 
{\cal L}_D=\alpha_D\phi_0^{2\frac{d+z(D-2)}{d-2+zD}}
\hat{R}^{(D)},
\nn\\
&&{\cal L }_F= -\frac{\alpha_F}{4}\phi_0^{2\frac{d+2+z(D-4)}{d-2+zD}}
\gamma^{mn}\gamma^{pq}g_{\mu\nu}F_{mp}^{\mu}F_{nq}^{\nu},\nn\\
 &&{\cal L}_K=-\frac{\alpha_K}{4}\phi_0^{2\frac{d+z(D-2)}{d-2+zD}}\gamma^{mn}g^{\mu\nu}g^{\alpha\beta}
 \left(\tilde{\cal
D}_m g_{\mu\alpha}\tilde{\cal D}_{n}g_{\nu\beta}
 -\alpha\tilde{\cal D}_m g_{\mu\nu}
 \tilde{\cal D}_{n}g_{\alpha\beta}\right),\nn\\
&&{\cal L}_R=-\frac{\alpha_R}{4}\phi_0^{2}
g^{\mu\nu}\gamma^{mn}\gamma^{pq}\left(
\tilde\nabla_{\mu}
\gamma_{mp}\tilde\nabla_{\nu}\gamma_{nq}
 -\beta\tilde\nabla_{\mu}
\gamma_{mn}\tilde\nabla_{\nu}\gamma_{pq}\right).\nn
\end{eqnarray}
If we assign an order one value to $\phi_0$, the coefficients, $\phi_0^2$ in front of $R^{(d)}$ in ${\cal L}_d$ and in ${\cal L}_R$ are still order unity respectively. However, $\phi_0$-dependence of ${\cal L}_D$, ${\cal L}_F$ and ${\cal L}_K$ are given by
\begin{eqnarray}
{\cal L}_D \sim {\cal L}_K \sim \exp\left({2\frac{d+z(D-2)}{d-2+zD}}\log \phi_0\right), \\
{\cal L}_F \sim \exp\left({2\frac{d+2+z(D-4)}{d-2+zD}}\log \phi_0\right).
\end{eqnarray}
These factors will suppress the Lagrangian density ${\cal L}_D$, ${\cal L}_F$ and ${\cal L}_K$ highly by the similarly mild adjustment that we discuss above.
For instance, when $d=D=2$, ${\cal L}_D $, ${\cal L}_F$ and ${\cal L}_K$ will be suppressed by a factor, $\sim e^{-a}$, where $a\geq 20$ for 
 $\phi_0\sim e=2.71828...$
and  $-0.1\leq z< 0 $.

\section{Lower-Dimensional Gravity}
\label{Lower Dimensional Gravity}

In this section, we   discuss several aspects of the anisotropic gravity theory (\ref{taction}) in the case with $d\leq3$.   
First, we  restrict to the case where  the coefficients $\alpha_D=\alpha_F=\alpha_K=\alpha_R=1 $ in (\ref{taction}) so that gauge-fixed $\phi=$ constant action  in the isotropic case $z=1$ with $\alpha=\beta=1$ reduces to the  Einstein-Hilbert action. Next, we introduce a scale $M_*$ which sets the scale for the conformal  symmetry breaking of the vacuum solution.
We  assume that  the field $\phi$ and the coordinate $y$  recover their canonical dimensions with the vacuum solution, and therefore we re-scale them via (\ref{res11})
%\begin{equation}
%\rightarrow M_*^{-z+1}y, ~~\phi\rightarrow M_*^{\frac{d-2+zD}{2}}\phi,
%\end{equation}
after which the coordinate $y$ has inverse-mass dimension  and $\phi$ becomes  a dimensionless field.
 
% \subsection{Lower Dimensional Gravity}
Let us first consider the case of $d=2.$ 
We notice that the ${\cal L}_D$, ${\cal L}_F$, ${\cal L}_K$ and the term of the scalar potential in  ${\cal L}_{d=2}$(being proportional to $V_0$) in (\ref{taction})  can become highly suppressed for small value of $z$ as was discussed in Sec. \ref{Suppression of the Lagrangian density}.  In fact, they become vanishing  in the limit  $z\rightarrow0^-$ together with  $\phi_0>1$
(or $z\rightarrow0^+$ with  $\phi_0<1$) , where the $\phi_0$ is the vev of the rescaled field $\phi$ as $\phi=\phi_0 M_*^{\frac{d-2+zD}{2}}$ via  (\ref{res11}).
%We observe that ${\cal L}_D$. ${\cal L}_F$, ${\cal L}_K$ and $V_0$ terms in   (\ref{taction}) vanished in the limit $z\rightarrow 0$ if $\phi$ is gauge-fixed to be  $\phi=\phi_0M_*^{\frac{d-2+zD}{2}} ~(\phi_0<1)$ using the conformal invariance. Furthermore, 
%\paragraph{2-$d$ non-linear $\sigma$-model}
The remaining ${\cal L}_{d=2}$ and ${\cal L}_R$ terms do not contain any $y$-derivatives and do not yield towers of massive modes. This corresponds to  maximally anisotropic  spacetime  and its dimension is reduced effectively to two. The  effective two-dimensional gravity theory after integrating over $y$ can be written as a two-dimensional conformal gravity theory coupled with matter 
\begin{equation}
S_{(2, D)}=\phi_0^2\Big(\frac{M_*}{M_e}\Big)^D\int d^2x \sqrt{-g^{(2)}}\Big[f(X)R^{(2)}
-g^{\mu\nu}G_{IJ}(X)\partial_\mu X^I\partial_\nu X^J\Big],\label{twodqg}
\end{equation}
where $M_e$ is the length associated with scale of the extra dimensions' coordinate volume, $[\int dy]^D\equiv M_e^{-D}$. The matter-looking like fields $X^I~(I=1,\cdots, (d+D-1)(d+D-2)/2$ are furnished by the metric $\gamma_{mn}$ by $\gamma_{33}=X^1, \gamma_{34}(=\gamma_{43})=X^2, \dots, \gamma_{d+D d+D}=X^{(d+D-1)(d+D-2)/2}$. The factor in front of the curvature scalar $R^{(2)}$ comes from $\sqrt{\gamma}$. The precise content of $f(X)$ and $G_{IJ}(X)$ will be given in a more familiar expression shortly.  $S_{(2,D)}$ can be regarded as two-dimensional gravity coupled with nonlinear sigma model of scalar matter fields. 

 Einstein gravity in $2+\epsilon$ spacetime dimensions and this type of extension has been studied extensively  and the most interesting aspect  is its  renormalizabilty \cite{Kawai:1995ju}. This aspect may have some application in the quantum theory of  (2,D) gravity theory. For $D=2$, one can show that action (\ref{twodqg}) can be put into %\newpage
\begin{eqnarray}
S_{(2, 2)}=\phi_0^2\Big(\frac{M_*}{M_e}\Big)^2\int d^2x \sqrt{-g^{(2)}}\Big[\varphi \Big(R^{(2)}-\frac{1}{4}g^{\mu\nu}\tilde\gamma^{mp}\tilde\gamma^{nq}\nabla_\mu\tilde\gamma_{mn}
\nabla_\nu\tilde\gamma_{pq}\Big)\nn \\ \nonumber
\end{eqnarray}
\vspace{-1.5cm}
\begin{eqnarray}
~~~~~~~~~~~~~~~~~~~~~~~~~~~~~~~~~~~~~-\frac{1}{2}(1-2\beta)g^{\mu\nu}{\varphi}^{-1}\partial_\mu \varphi \partial_\nu \varphi \Big],\label{xtwodqg}
\end{eqnarray}
where $\varphi\equiv\sqrt{\gamma}$ and $\tilde\gamma_{mn}=\gamma_{mn}/\sqrt{\gamma}$ with $\sqrt{\tilde\gamma}=1.$ 
For $2\beta\leq 1$, this theory is ghost free. 

One may consider (\ref{xtwodqg}) with $\beta\leq1/2$ as a UV-completion of Einstein gravity.
The coupling 
$\beta$ and the parameter $z$ are amenable to  scale-dependency and may show $z=1$ and $\beta=1$ fixed point in IR, where the theory becomes Einstein theory. This aspect of the lower-dimensional anisotropic gravity is one of the novel features in relation with four-dimensional quantum gravity and it deserves to be studied further.
%If it  could flow to the Einstein gravity with $z=1$ and $\beta=1$ by assuming $z$-dependence of the coefficient $\beta,$ it has potential to be UV completion of quantum gravity; The high behaviour is governed by two-dimensional $z=0$ theory and at low energy the theory flows to isotropic  $z=1$.
%This deserves further study.

When $D=1,$  ${\cal L}_D$ and ${\cal L}_F$ terms vanish identically in (\ref{taction}). For $\beta\neq 1$, ${\cal L}_R$ contains  a single scalar given by $\gamma_{d+1 d+1}$. ${\cal L}_K$ term which involves dynamics along $y$-direction  can again be made to vanish with $z\rightarrow2-d+0^-$ and $\phi_0>1$. The resulting effective $d$-dimensional theory is gravity coupled with a single scalar field $\gamma_{d+1 d+1}\equiv N^2(x)$.
%But in the latter case, there is non-vanishing cosmological constant term %given by $V_0\phi_0^{\frac{2}{d-1}}M_*^2$. 
More explicitly, for $d=3$ and $z=-1$ we have 
\begin{equation}
S_{(3, 1)}=\phi_0^2\Big(\frac{M_*^2}{M_e}\Big)\int d^3x \sqrt{-g^{(3)}}\Big[R^{(3)}-
\frac{1}{2}(\frac{3}{2}-\beta)g^{\mu\nu}\partial_\mu \varphi \partial_\nu \varphi \Big].\label{xthreeqg}
\end{equation}
 In the above action, we rescaled $g_{\mu\nu}\rightarrow N^{-1}g_{\mu\nu}$ and redefined $\varphi\equiv\sqrt{2}\log N$. For $\beta<3/2$ it is ghost-free and  and  this demonstrates that $d=3$ anisotropic gravity  furnishes alternative perspective to the three-dimensional gravity.
   
%\textcolor{blue}
{Another interesting limit  in $(d,D)=(2,1)$ appears when take a limit that $z\rightarrow0^-$ together with  $\phi_0>1$ and $V_0\rightarrow\infty$ by keeping 
that $V_0\phi_0^{\frac{d+zD}{d-2+zD}}\equiv  2 \bar\Lambda$ is held fixed. In this case, the effective 2-dimensional action is given by
\begin{equation}
S^\Lambda_{(2,1)}=\phi^2_0\frac{M_*}{M_e}\int d^2 x e^{-2\psi}\left[R^{(2)} -8(1-\beta)g^{\mu\nu}\partial_\mu\psi\partial_\nu\psi-2\bar \Lambda\right],
\end{equation}
where $\sqrt{\gamma}=\sqrt{\gamma_{d+1,d+1}}\equiv e^{-2\psi}$. When we choose that the $
\beta=\frac{1}{2}$, then this action becomes the CGHS model action without matters\cite{Callan:1992rs}.

\section{Vacuum Solution}
\label{Vacuum Solution}
In this section, we search  for the solutions of equation of motion derived from (\ref{taction}). We start from a simple ansatz which solves them instead of writing down the  tedious looking equations \footnote{See Ref. \cite{Moon:2017rox} in some special case with $d=4$ and $D=1$}.  For our purpose, the detailed equations of motion
are not necessary.  Consider the following ansatz
\begin{equation}
g_{\mu\nu} \equiv g_{\mu\nu}(x), ~ \gamma_{mn}\equiv \gamma_{mn}(y),~\phi=\phi_0,~ N^\mu_m=0,
\label{ansatz}
\end{equation}
which yields $F_{mp}^{\mu}=\tilde{\cal D}_m g_{\mu\alpha}=\tilde\nabla_{\mu}
\gamma_{mp}=0$.
  
%unless $z=1$ so that $v(z)=1+D/2$, i.e., the isotropic case.
% From here on, we
% choose $v=(2+Dz)/2$, because it simply yields the conventional 4D
%non-minimal coupling term of $\phi^2 R^{(4)$. The case $Dz=-2$ is given %a separate treatment.}
 
Since ${\cal L}_F, {\cal L}_K,$ and ${\cal L}_R$ are quadratic in $F_{mp}^{\mu}, \tilde{\cal D}_m g_{\mu\alpha}$, and $\tilde\nabla_{\mu}
\gamma_{mp}$ respectively,  the ansatz (\ref{ansatz}) 
will automatically solve the equations of motion derived from these terms.
Then, effectively one can consider 
\begin{eqnarray}
S^\prime=M_*^{d-2+D}\int d^{d}xd^{D}y \sqrt{-g}\sqrt{\gamma}
\Bigg[\phi^2R^{(d)}
 +\phi^{2\frac{d+z(D-2)}{d-2+zD}}
 \hat{R}^{(D)}-V_0M_*^2\phi^{2\frac{d+zD}{d-2+zD}}\Bigg].\label{effaction}
\end{eqnarray}
Note that the over-all gravitational  scale $M_*$ appears as a consequence of symmetry breaking in (\ref{effaction}).
The equations of motion are derived as follows:
\begin{eqnarray}
\phi^2\left(R^{(d)}_{\mu\nu}-\frac{1}{2}g_{\mu\nu}R^{(d)}\right)&=&
-\frac{1}{2}V_0M_*^2\phi^{2\frac{d+zD}{d-2+zD}}g_{\mu\nu}+
\frac{1}{2}\phi^{2\frac{d+z(D-2)}{d-2+zD}}
 \hat{R}^{(D)}g_{\mu\nu},\nn\\
\phi^{2\frac{d+z(D-2)}{d-2+zD}}\left(\hat R^{(D)}_{mn}-\frac{1}{2}\gamma_{mn}\hat R^{(D)}\right)&=&
-\frac{1}{2}V_0M_*^2\phi^{2\frac{d+zD}{d-2+zD}}\gamma_{mn}+
\frac{1}{2}\phi^{2}\hat{R}^{(d)}\gamma_{mn},\nn\\ 
\phi^2R^{(d)}+\tiny{\frac{d+z(D-2)}{d-2+zD}}\phi^{2\frac{d+z(D-2)}{d-2+zD}}
 \hat{R}^{(D)}&=&V_0\frac{d+zD}{d-2+zD}M_*^2\phi^{2\frac{d+zD}{d-2+zD}},
\end{eqnarray}
where the terms containing derivatives of $\phi$ is all ignored since we assume that $\phi=\phi_0$. To solve the above equations, we make further ansatz
\begin{eqnarray}
R^{(d)}_{\mu\nu}=\Lambda^{(d)}g_{\mu\nu}(x), ~ 
~R^{(D)}_{mn}=\Lambda^{(D)}\gamma_{mn}(y).
\end{eqnarray}
Then, one obtains the following values of $\Lambda^{(d)}$ and $\Lambda^{(D)}$:
\begin{eqnarray}
\Lambda^{(d)}=\frac{V_0}{d-2+D}\phi_0^{\frac{4}{d-2+zD}}M_*^2,~
\Lambda^{(D)}=\frac{V_0}{d-2+D}\phi_0^{\frac{4z}{d-2+zD}}
M_*^2.~~~~(D\geq 2)\label{sizesep}
\end{eqnarray}

From (\ref{sizesep}), $d$-dimensional spacetime can be characterized by a positive cosmological constant if  $V_0>0$ (de Sitter) whereas a negative one if $V_0<0$
(anti-de sitter).  The sign of $\Lambda^{(D)}$
determines whether the extra dimensions are compact ($\Lambda^{(D)}>0)$ or
non-compact  ($\Lambda^{(D)}<0)$.
Therefore, we have two possibilities
\footnote{For a general value of $\alpha_D$ different from 1, we have $\Lambda^{(D)}=\alpha_D^{-1}\frac{V_0}{d-2+D}\phi_0^{\frac{4z}{d-2+zD}}
M_*^2$. Then, we have two more possibilities; (iii)  de Sitter$(d)\times NC^{D}$ for $V_0>0, \alpha_D<0$ and (iv) anti-de Sitter$(d)\times C^{D}$ for $V_0<0, \alpha_D<0.$\\}:
\\
(i) de Sitter$(d)\times C^{D}$ for $V_0>0, \alpha_D>0.$\\
(ii) anti-de Sitter$(d)\times NC^{D}$ for $V_0<0, \alpha_D>0.$
\\
Here, $C^D(NC^D)$ denotes compact (non-compact) $D$-dimensional extra space. 

The ratio of $\Lambda^{(d)}$ to $\Lambda^{(D)}$ is given by
\begin{equation}
\Lambda^{(d)}=\phi_0^{4\frac{1-z}{d-2+zD}}\Lambda^{(D)}.
\end{equation}
When $z=1$, we have $\Lambda^{(d)}=\Lambda^{(D)}$, and there is no size separation between $d$-dimensional spacetime and $D$-dimensional extra space.  Large separation can occur when $z$ is again near the value $(2-d)/D.$ For $d=4$ and $z=\frac{-2}{D}+ \epsilon$ we have $\phi_0^{4\frac{1-z}{d-2+zD}}= \exp\left[ 4\left(\frac{2+D}{\epsilon D^2}-\frac{1}{D}\right)\log\phi_0\right]$ and it can become very huge  when $\epsilon$ is a small(but not very much small) and positive number with order one value of $\phi_0<1$(or $\epsilon$ is negative number with order one value $\phi_0>1$).

For a specific example, let us consider $d=4$ and $D=2$ case.
The ratio of the cosmological constants $\Lambda^{(d=4)}$ and $\Lambda^{(D=2)}$ 
%which appear in the vacuum solution %defined in $4-$ and extra $1-$dimensions 
is given by
\begin{equation}
\Lambda^{(d=4)}\sim 10^{-124}\Lambda^{(D=2)},
\end{equation}
%and has no
%For example, when $z=\frac{2-d}{D}+\epsilon$, where $\epsilon$ is rather small and positive number, then $\phi_0^{4\frac{1-z}{d-2+zD}}=e^{ 4\frac{1-z}{d-2+zD}\log \phi_0}$
% becomes very much small by assuming that $\phi_0\sim 10$. If $\epsilon\sim-0.1$, then  
%\Lambda^{(d)}\sim 10^{-124}\Lambda^{(D)}$.}
when $\phi_0\sim 10$ and $\epsilon\sim-0.03$.
We note that if $\Lambda^{(2)}$ is the Planck scale, $\Lambda^{(4)}$ comes close to the current value of the cosmological constant in our universe.
%When $z=1$ and $\alpha_D=1,$ we have $\Lambda^{(d)}=\Lambda^{(D)}$, i.e., %isotropic case.
%\textcolor{red}{We take $\phi_0\sim 10.$ : question?}

%-----$D=1$ sol-----

%\noindent Let us discuss large extra dimension.
{Now let us discuss the relation between Planck length and the size of the extra dimensions}. We define the $d$-dimensional Planck length $l_P$ from (\ref{effaction}) (with
$l_F=M_*^{-1}$)
\begin{eqnarray}
l_P=\left[\phi_0^2l_e^Dl_F^{-d+2-D}\right]^{\frac{1}{-d+2}}~~(D\geq1)\label{planck_const}
\end{eqnarray}
with Newton constant $G_F=l_P^2.$ Then, We have size of the extra dimensions given by
\begin{equation}
l_e=l_F\phi_0^{-\frac{2}{D}}\left(\frac{l_F}{l_P}\right)^{\frac{d-2}{D}}.
\label{scalesep}
\end{equation} 

%\textcolor{blue}
{We find that the expression (\ref{scalesep}) has nothing to do with the anisotropic factor $z$. 
This expression would be worthy of attention in a sense that it is very similar to the equation (4) given in \cite{Arkani-Hamed:1998jmv}. In their model in \cite{Arkani-Hamed:1998jmv}, the main result of condition on the size of extra dimension to get electro-weak scale from Planck scale(to resolve hierarchy problem) is 
\begin{equation}
l_E\sim 10^{30/n-17}\left(\frac{1\rm{TeV}}{m_{\rm{EW}}} \right)\rm{cm},
\end{equation}
where $l_E$ is the size of the extra $n$-dimension, and $m_{\rm{EW}}$ is the electro-weak scale. When $n=1$, $l_E$ becomes $10^{13} \rm{cm}$ which is too large to accept. However, $n\geq 2$, the value of $l_E$ becomes more reasonable. For instance, if $n=2$, then $l_E$ is around $10^{-3}\rm{cm}$.

{One may say that the size of the extra dimension as $l_e=10^{-3}$cm is still enough to be detected by experiment. However, in string theory, by assuming that we are living on a D3-brane, only closed string can probe the extra dimensions, whose massless excitations correspond to gravitational fields. It is not certain whether gravity obeys inverse-square law below or near about $10^{-2}$cm yet.  If the gravity force becomes weaker below this length scale and becomes $\sim r^{-k}$ with $k>2$, then that could be an indication of extra dimensions.}
Of course, $l_E=10^{-3}$cm in dimensions other than $d=4$ is not well motivated by experiment.

%We note that identical to the one given in the brane-world models
%except the factor $\phi_0$ necessary for conformal invariance\cite{Arkani-Hamed:1998jmv}.   
%Another crucial difference is that  $M_*$ which we introduced for anisotropic conformal symmetry breaking cannot be considered as $(d+D)$-dimensional Planck scale; we do not have the notion of $(d+D)$-dimensional Planck length (at least classically), because the theory is intrinsically formulated in anisotropic spacetime. 
%But we have $d$-dimensional Newton constant.
From the relation (\ref{scalesep}), we get a similar result in \cite{Arkani-Hamed:1998jmv}.
We assume $l_F\sim 10^{-18}$ cm which is somewhat lower than the present-day accelerators can explore.
%(\ref{scalesep}) displays many possibilities including large extra dimensions. 
When $d=4$ and $D=1$, with $l_{P}=10^{-33}$cm, we get $l_e\sim 10^{12}$cm by assuming that $\phi_0\sim 1$. This is too large to accept too. %However, for $\phi_0\sim 10^{7.5}$, we can have $l_e\sim 10^{-3}$cm.  
However, when $d-2=D$ which includes $d=4$ and $D=2$ case, we have a coincidence that  the size of the extra dimensions
is always given by $l_e=10^{-3}$cm 
for $\phi_0\sim 1$\cite{Arkani-Hamed:1998jmv,Zwiebach:2004tj}. }

%\textcolor{blue}{Note expression (\ref{scalesep}) is independent on anisotropic factor $z$ and almost identical to the one given in the brane-world models
%except the factor $\phi_0$ necessary for conformal invariance\cite{Arkani-Hamed:1998jmv}. Another crucial difference is that  $M_*$ which we introduced for anisotropic conformal symmetry breaking cannot be considered as $(d+D)$-dimensional Planck scale; we do not have the notion of $(d+D)$-dimensional Planck length (at least classically), because the theory is intrinsically formulated in anisotropic spacetime. 
%But we have $d$-dimensional Newton constant.
%We assume $l_F\sim 10^{-18}$ cm which is somewhat lower than the present-day accelerators can explore.
%(\ref{scalesep}) displays many possibilities including large extra dimensions. 
%When $d=4$ and $D=1$, $l_{p}=10^{-33}$cm, $l_e\sim 10^{12}$cm for $\phi_0\sim 1.$ However, for $\phi_0\sim 10^{7.5}$, we can have $l_e\sim 10^{-3}$cm.  
%When $d-2=D$, we have a strange coincidence that  the size of the extra dimensions
%is always given by $l_e=10^{-3}$cm for $\phi_0\sim 1$\cite{Zwiebach:2004tj,Arkani-Hamed:1998jmv}. }

%\textcolor{blue}

%more example; $d=4, D=2$,  $d=4, D=6.$ }

%\textcolor{blue}{We note that the second kind solution, anti-de Sitter$(d)\times NC^{D}$ can be obtained from supergravity theories inspired by string and M- theories too. From a certain gauged supergravity theories, AdS$_d\times R^2$ spacetimes are constructed\cite{Donos:2011pn,Almuhairi:2011ws},where $d=2 \rm \ or\ 3$.}
\section{Discussions}

We investigated various aspects of anisotropic gravity in which the scaling properties of the $d$-dimensional spacetime and the extra $D$ dimensions are different %treated differently with the introduction of anisotropic factor 
and characterized by the parameter $z$. 
%It is exhibited that $z$ can be effective in explaining some issues in $(d+D)$-dimensional gravity. 
The parameter $z$ is a crucial factor in our model.
The most salient feature is that $z$ can act as an agent to (de)activate the extra dimensions for some specific value and induce the effective dimensional reduction. When $z=-\frac{d-2}{D},$ the kinematics along the extra dimensions are completely suppressed and the extra dimensions are virtually obsolete. 
This %could have applications in 
suggests a quantum gravity where
%by reducing 
the four-dimensional Einstein gravity reduces to a certain two-dimension theory:  Einstein gravity starts effectively in two-dimensional spacetime at $z=0$ in UV where it is renormalizable and ghost-free, and it flows to isotropic $z=1$ in IR recovering the four-dimensional spacetime. It also provide a possibility to address the hierarchy problem and the origin of  large scale separation between the $d$-dimensional spacetime and the extra $D$ dimension in physics.

\section*{Appendices}
\subsection{Matter field action}
We consider the following matter field action which is invariant under the FPD and the scale transformation. 
\begin{equation}
S_{matter}=\int d^dxd^Dy\sqrt{-g}\sqrt{\gamma}\left[ \frac{1}{2}{\phi^2}g^{\mu\nu}\partial_\mu \Phi \partial_\nu \Phi
+\frac{1}{2}\phi^{2\frac{d+zD-2z}{d-2+zD}}\gamma^{mn}\partial_m \Phi \partial_n \Phi
-\phi^{\frac{2(d+Dz)}{d-2+Dz}}f(\Phi)\right],
\end{equation}
where the $f(\Phi)$ is an arbitrary function of the matter field $\Phi$. As we discussed, the field $\phi$ has a vev, $\phi_0$. With a rescaling as $\Phi=(\sqrt{V_D}\phi_0)^{-1}\tilde\Phi$, the action becomes
\begin{equation}
\label{matter-action2}
S_{matter}=\int d^dx\sqrt{-g}\int\frac{ d^Dy\sqrt{\gamma}}{V_D}\left[\frac{1}{2}g^{\mu\nu}\partial_\mu \tilde\Phi \partial_\nu \tilde\Phi
+\frac{1}{2}\phi^{\frac{4(1-z)}{d-2+zD}}\gamma^{mn}\partial_m \tilde\Phi \partial_n \tilde\Phi
-\mathcal V_{eff}\right].
\end{equation}
What we want to get is an effective $d$-dimensional action from this by supposing that the matter field $\tilde\Phi$ does not depend on the coordinate of extra dimensions, $y^m$, i.e. $\tilde \Phi\rightarrow \tilde \Phi(x^\mu)$. This causes that the second term in (\ref{matter-action2}) vanishes and it lets the factor, $\int\frac{ d^Dy\sqrt{\gamma}}{V_D}\rightarrow 1$.
The effective potential $\mathcal V_{eff}$ is given by
\begin{eqnarray}
\label{eff-pot}
\mathcal V_{eff}&=&V_DM_*^{D+d}\phi_0^{\frac{2(d+Dz)}{d-2+Dz}}f\left(\frac{M_*^{\frac{2-D-d}{2}}\tilde\Phi}{\sqrt{V_D}\phi_0}\right) \\ \nonumber
&\rightarrow&\sum_{n=1}^\infty \frac{a_n}{n!}\left( V_DM_*^{D+d}\phi_0^2\right)^{1-\frac{n}{2}}
M_*^n \phi_0^{\frac{4}{d+zD-2}}\tilde \Phi^n.
\end{eqnarray}
For the last line in (\ref{eff-pot}), we assume that the $f(X)=\sum_{n=2}^\infty \frac{a_n}{n!}X^n$ with constants, $a_n$.
Recall (\ref{planck_const}) and (\ref{scalesep}) then we realize that $V_DM_*^{D+d}\phi_0^2=M_*^2M_p^{d-2}$. Therefore, the effective potential becomes
\begin{equation}
\mathcal V_{eff}=\sum_{n=1}^\infty \frac{a_n}{n!}\left( M_*^2M_p^{d-2}\right)^{1-\frac{n}{2}}
M_*^n \phi_0^{\frac{4}{d+zD-2}}\tilde \Phi^n.
\end{equation}
The effective potential depends on the $d$-dimensional Planck energy scale, $M_*$ and the $z$. In fact, it relies on the $D$-dimensional volume implicitly since $M_p$ is determined once $M_*$, $\phi_0$ and $l_e$ are to be chosen(See equations, (\ref{planck_const}) and (\ref{scalesep})). %The radial directional perturbative field $\delta \Phi$ around the 

\subsection{Higgs vev}
One may truncate the effective potential up to an order of $\tilde \Phi^4$ and demand that $a_2=-|a_2|$, $a_3=0$ and $a_4$ is positive to form a Mexican-hat potential. The spontaneous symmetry breaking occurs at $\frac{\partial \mathcal V_{eff}}{\partial \tilde \Phi}=0$, where the field $\tilde \Phi$ has its vev in the real vacuum in the presence of such a potential. The value of the vev is given by
\begin{equation}
\Phi^2_0=-3!\left(\frac{a_2}{a_4}\right)M_p^{d-2}.
\end{equation}
The radial directional perturbation around the vacuum in the field space of $\tilde \Phi$ produces a massive particle and its physical mass also shows a large scale difference from the scale $M_*$ as $M_{phys}=\exp\left({\frac{2}{d-2+Dz}\log\phi_0}\right)$ when  $z=\frac{2-d}{D}\pm \epsilon$ where $\epsilon$ is a small(but not very much small) and positive number with $\phi_0 <1$.

~~\\~~\\
\noindent {\bf Acknowledgement}

J.H.O would like to thank his $\mathcal W.J.$ and $\mathcal Y.J.$ This work was supported by the National Research Foundation of Korea(NRF) grant funded by the Korea government(MSIP) (No.2016R1C1B1010107). This work is also partially supported by Research Institute for Natural Sciences, Hanyang University.

%%%%%%%%%%%%%%%%%%%%%%%%%%%%%%%%%%%%%%%%%%%%%%%%%%%%%%%%%%%%%%%%%%%

\end{document}